\documentclass[aps,preprint]{revtex4}%
\usepackage{amsfonts}
\usepackage{amsmath}
\usepackage{amssymb}
\usepackage{graphicx}%
\begin{document}
\preprint{ }
\title[ ]{The Paradoxical Forces for the Classical Electromagnetic Lag Associated with
the Aharonov-Bohm Phase Shift}
\author{Timothy H. Boyer}
\affiliation{Department of Physics, City College of the City University of New York, New
York, New York 10031}

\begin{abstract}
The classical electromagnetic lag associated with the Aharonov-Bohm phase
shift is obtained by using a Darwin-Lagrangian analysis similar to that given
by Coleman and Van Vleck to identify the puzzling forces of the Shockley-James
paradox. \ The classical forces cause changes in particle velocities and so
produce a relative lag leading to the same phase shift as predicted by
Aharonov and Bohm and observed in experiments. \ \ An experiment is proposed
to test for this lag aspect implied by the classical analysis but not present
in the currently-accepted quantum topological description of the phase shift.

\end{abstract}
\maketitle

\section{Introduction}

In the past 45 years, the Aharonov-Bohm phase shift\cite{AB} has passed from a
surprising novelty into a standard part of the quantum mechanical textbook and
research literature.\cite{G-Q} \ Yet the nature of the effect may well be misunderstood.

Aharonov and Bohm\cite{AB} attracted attention to the phase shift, which had
been suggested earlier by Ehrenberg and Siday,\cite{E-S} by claiming that the
effect involved a new role for the electromagnetic potentials in quantum as
compared to classical physics, and that the phase shift arose as a quantum
topological effect occurring in the absence of classical electromagnetic
forces. \ A contrary view has also been urged.\cite{B} \ It has been suggested
that the experimentally observed phase shift results from a lag effect
involving classical electromagnetic forces. \ Although the Aharonov-Bohm phase
shift itself has been well verified experimentally, there is no experimental
evidence indicating whether or not the effect involves classical forces and
velocity changes.\cite{M-P} \ The support for the quantum topological nature
of the shift comes only from theoretical arguments which claim that classical
electromagnetic forces can not possibly cause the phase shift. \ However, it
has been suggested repeatedly that electromagnetic theory actually supports
exactly the contrary view. \ Now at last, the forces involved are identified
in some detail in the discussion of the following paper.\cite{Bill} \ The
observed phase shift can indeed be accounted for by a semiclassical analysis
based upon a classical electromagnetic lag effect. \ Therefore it is now time
for further experiments measuring aspects of the phase shift which differ
between the two alternative explanations. \ 

\section{Force Paradoxes}

The basic classical electromagnetic interaction involved in the Aharonov-Bohm
phase shift involves a point charge and a magnet. \ Such an interaction was
discussed in 1968 in an often-quoted paper by Coleman and Van Vleck,\cite{CVV}
"Origin of 'Hidden Momentum Forces' on Magnets." \ The authors used the Darwin
Lagrangian to treat the interaction to order $1/c^{2}$. \ Their discussion
centers on the paradox of Shockley and James\cite{S-J} involving a changing
magnetic moment which leads to an easily identifiable force on an external
charge, and yet there seems to be no force back on the magnet. \ Momentum
conservation seems to fail. \ The situation discussed by Coleman and Van Vleck
seems remarkably similar to the situation of the Aharonov-Bohm phase shift
involving charged particles passing a magnetic solenoid or toroid. \ Because a
moving charged particle causes a magnetic field, it puts an easily
identifiable Lorentz force on the currents of the magnet, and yet for a toroid
or long solenoid there seems to be no force back on the passing charge. \ In
both cases, the same fundamental interaction between a charged particle and a
magnet is involved. \ Indeed, Coleman and Van Vleck have provided some aspects
needed to understand the classical electromagnetic lag connected with the
Aharonov-Bohm phase shift.

In this article, we point out the classical electromagnetic energy and
momentum changes which must be accounted for in understanding the classical
electromagnetic aspects of the Aharonov-Bohm situation. \ In the following
article,\cite{Bill} we turn to a classical hydrogen atom as a model of a
magnetic moment and treat its interaction with a passing charge using the
Darwin Lagrangian, illustrating in detail aspects of the Coleman-Van Vleck
analysis and the forces associated with the energy and momentum changes; then
we comment on the passage to a many-particle situation for the case of a
magnetic toroid configuration.

\section{Use of the Darwin-Lagrangian Analysis}

The Aharonov-Bohm situation involves a point charge moving past a magnet in
the form of a long solenoid or toroid. \ Let us imagine that external forces
are applied to the charges carrying the currents of the magnet and also to the
passing charge $q$ so that the non-interacting motions are preserved despite
the interaction of the magnet and the charge due to their electromagnetic
fields. \ As shown previously,\cite{B73} the net work $W_{ext}$ done by the
external forces provides the energy of overlap of the magnetic field of the
passing charge and the magnet, and also the net impulse $I_{ext}$ delivered by
the external forces accounts for the change in linear momentum in the
electromagnetic fields. \ These relations take the form%
\begin{equation}
W_{ext}=(q/c)\mathbf{v}_{q}\cdot\mathbf{A}_{\mu}(\mathbf{r}_{q}(t))
\end{equation}%
\begin{equation}
\mathbf{I}_{ext}=(q/c)\mathbf{A}_{\mu}(\mathbf{r}_{q}(t))
\end{equation}
where $\mathbf{A}_{\mu}(\mathbf{r}_{q}(t))$ is the vector potential in the
Coulomb gauge of the magnet evaluated at the position $\mathbf{r}_{q}(t)$ of
the passing charge. \ Thus the Aharonov-Bohm situation involves energy and
momentum changes which can be associated with the vector potential of the magnet.

Of course, an experimental situation does not involve external forces which
maintain the currents constant and the distant particle at constant velocity.
\ Thus the crucial question is just how do \ the currents and particle
velocity change so as to give energy and momentum conservation and also
constant motion of the system center of energy. \ In the following article, we
analyze this problem in a simple model for a magnetic moment, and then discuss
the multiparticle limit to form a toroid. We suggest that the charge and
current densities of the magnet change in such a way as to screen the passing
electric field out of the magnet and to produce an electric field back at the
passing charge which changes the velocity of the passing charge as%
\begin{equation}
m\Delta\mathbf{v}_{q}=(q/c)\mathbf{A}_{\mu}(\mathbf{r}_{q}(t))
\end{equation}
As shown in the following article, these changes are consistent with ideas of
conservation of energy, of linear momentum, and for constant motion of the
center of energy. \ Furthermore, the net Lorentz forces between the toroid and
passing charge satisfy Newton's third law.

\section{Calculation of the Aharonov-Bohm Phase Shift}

The result in Eq. (3) is also exactly what is required to account for the
Aharonov-Bohm phase shift as a classical electromagnetic lag effect. \ A point
charge moving with initial velocity $\mathbf{v}_{0}$ down the axis of a toroid
will suffer a relative displacement compared to a point charge which passes
outside the toroid and experiences no forces. \ The relative displacement in
the direction of motion when the point charge passes through the toroid can be
found by integrating the magnitude of the velocity change $\Delta\mathbf{v}$
in Eq. (3), taken here as in the $y$-direction,
\begin{equation}
\Delta Y=%
{\displaystyle\int\limits_{-\infty}^{\infty}}
dt\,\Delta v_{y}=%
{\displaystyle\int\limits_{-\infty}^{\infty}}
\frac{dy}{v_{0}}\,\frac{q}{mc}A_{y}(\mathbf{r})
\end{equation}
The semiclassical phase shift $\Delta\phi=p_{y}\Delta Y/\hbar=mv_{y}\Delta
Y/\hbar$ associated with this relative lag is therefore%
\begin{equation}
\Delta\phi=\frac{1}{\hbar}%
{\displaystyle\int\limits_{-\infty}^{\infty}}
dy\,\frac{q}{c}A_{y}(\mathbf{r})=\frac{q\Phi}{\hbar c}%
\end{equation}
where $\Phi$ is the magnetic flux enclosed in a path around the toroid. \ Here
we have used $%
{\displaystyle\oint}
\mathbf{A}\cdot d\mathbf{r}=\Phi$ and can imagine closing the path of
integration along a curve which passes outside and far from the toroid. \ The
same phase shift in Eq. (6) was predicted by Aharonov and Bohm,\cite{AB} and
is observed experimentally.\cite{Chamb} \ 

\section{Experimental Tests for Velocity Changes}

The Aharonov-Bohm phase shift involves the shift of the double-slit electron
interference pattern inside an undisplaced single-slit pattern\cite{BB} as
shown in Fig. 1. \ The electron beams pass around a solenoid (or toroid), and
the shift of the double-slit pattern is proportional to the magnetic flux
enclosed. \ According to Aharonov and Bohm,\cite{AB} the phase shift is a
quantum topological effect and involves no forces on the passing electrons and
no velocity changes of the electrons. \ The alternative classical-based
explanation\cite{B} views the phase shift as analogous to placing a piece of
glass behind one slit of a double-slit optical interference experiment, with
the phase shift occurring because of a time lag between the transit times for
the two beams. \ This situation is shown in Fig. 2. \ In the classical-based
explanation of the Aharonov-Bohm phase shift, there is a relative lag between
the electrons passing through the two different slits based upon classical
forces which slow the electrons differently on the opposite sides of the
solenoid. \ An exactly analogous phase shift due to the electrostatic forces
of a line of electric dipoles was observed experimentally by Matteucci and
Pozzi\cite{MandP} and is sketched in Fig. 3. \ The electrostatic phase shift
clearly involves velocity changes due to electrostatic forces, analogous to
the situation of the optical phase shift in Fig. 2. \ It is now time for
experiments which do not simply verify the existence of the Aharonov-Bohm
phase shift, but rather which test whether or not the phase shift involves
velocity changes for the electrons. \ 

One possible experimental test looks for a breakdown of the double-slit
interference pattern at large magnetic flux. \ According to Aharonov and
Bohm's view, the magnetic flux in the solenoid can be made arbitrarily large
and the double-slit phase shift will never break down. \ According to the lag
view (again in analogy with classical optics where a wave passing through a
piece of glass suffers a velocity change), if the relative lag between the
beams becomes larger than the coherence length, then the double-slit
\ interference pattern will indeed break down. \ Testing for this break-down
of the interference pattern at large flux seems a feasible experiment; a large
magnetic flux can be inserted between the electron beams either by the use of
many solenoids or a solenoid of very large flux.

I know of no other aspect of optics or wave-aspects of particles where a phase
shift is claimed to occur without any other change whatsoever in the system.
\ If experiments show that there are no velocity changes involved in the
Aharonov-Bohm phase shift, then nature is indeed revealing an important
nonclassical aspect. \ If there are velocity changes corresponding to a lag
effect, as I believe is overwhelmingly likely, then the quantum topological
interpretation is untenable, and the phase shift is consistent with an
unfamiliar aspect of classical electromagnetic theory.

\section{Acknowledgment}

I wish to thank Mrs. Laurette B. Fisher for her assistance with the diagrams.

\section{Figure Captions}

Fig. 1. \ The Aharonov-Bohm Phase Shift. \ The double-slit interference
pattern of charged particles arriving at a distant screen is shifted by a long
solenoid (a line of magnetic dipoles) inserted between the beams. \ The
mechanism for the shift is in dispute. \ At present it is not known whether or
not the passing particles experience velocity changes associated with the
interference pattern shift.

Fig. 2. Optical Phase Shift Due to a Lag Effect. \ The double-slit
interference pattern on a distant screen is shifted when a piece of glass is
introduced behind one of the slits. \ The waves passing through the glass are
slowed down by the piece of glass.

Fig. 3. The Matteucci-Pozzi Phase Shift. \ The double-slit interference
pattern of charged particles arriving at a distant screen is shifted by a pair
of line charges (a line of electric dipoles) inserted between the beams. \ The
difference in the time order of the electrostatic forces experienced by the
charges passing through the two slits leads to an electrostatic lag effect
which accounts for the interference pattern shift.\cite{MandP}

\end{document}